\documentstyle[aps,preprint,epsfig]{revtex} 
\begin{document}                                     
\title{Occurence Probabilities of Stochastic Paths}
\author{Dirk Helbing and Rolf Molini}
\address{II. Institute for Theoretical Physics, University of
Stuttgart, 70550 Stuttgart, Germany}
\maketitle      
\begin{abstract}
An analytical formula for the occurence probability of Markovian stochastic
paths with repeatedly visited and/or equal departure rates is derived.
This formula is essential for an efficient investigation of the trajectories
belonging to random walk models 
and for a numerical evaluation of the
`contracted path integral solution' of the discrete master equation
[Phys. Lett. A {195}, 128 (1994)].
\end{abstract}
\section{Introduction}

Stochastic processes play an important role in all scientific fields dealing
with systems that are subject to inherent or external random influences
(`fluctuations'). Therefore, numerous methods and models have been developed
for the description of stochastically behaving systems 
\cite{Gard,Kamp,Haus,Hel95}.
The fields of application reach from physics \cite{Kamp}
over chemistry \cite{Kamp,Opp} and biology \cite{Lef} 
to the economic and social sciences \cite{Hel95,Soz}.
\par
In this paper we will focus on {\em Markovian} stochastic processes which
do not essentially depend on so-called {\em memory effects} so that
correlations of present transitions with past states of the
considered system (except the last one) can be neglected. We can distinguish 
{\em cross-section oriented methods} describing the temporal evolution 
of the {\em distribution of states} and {\em longitudinal-data 
oriented methods} 
delineating single stochastic {\em trajectories (time series)}. If one is 
confronted with a {\em continuous state space} the distribution of states
is governed by the {\em Fokker-Planck equation} \cite{Risk,Gard,Hel95}
and corresponding trajectories follow the {\em Langevin equation 
(stochastic differential equation)}
\cite{Risk,Gard,Hel95}. Related methods exist for quantum mechanical systems
\cite{quant}. In the following we will concentrate on systems
with a {\em discrete state space}. Then, the distribution of states
satisfies the {\em master equation} \cite{Gard,Hel95} 
whereas the corresponding
time series are determined by {\em random walk models} \cite{Haus}
and generated by means of {\em Monte-Carlo simulations} \cite{Bind}. 
\par
In a recent paper \cite{Hel94}, an important relation 
between the distribution of 
states and the occurence probabilities of paths %(sequences of system states)
has been established which was called the `contracted path-integral
solution'. However, during its numerical implementation, this relation turned
out to be restricted to the infrequent case of
pure {\em birth processes} (uni-directional transitions),
since it did not provide an analytical formula for
paths with repeatedly visited states. Therefore, this paper presents the
non-trivial derivation of the missing 
occurence-probability formula for `degenerate paths' (Sec. III). 
\par
The numerical implementation of the `contracted path-integral solution' is
outlined in Section IV. Since the {\em `breadth-first' 
procedure} \cite{Dep} %suggested in Ref. \cite{Hel94}
is very inefficient with respect to computer time and memory
(even for a few system states only), our algorithm
bases on the {\em `depth-first' procedure} \cite{Dep}. 
The suitability and correctness of
this new numerical method is illustrated by an example concerning
Brownian motion. 
\par
Section V summarizes the results of the paper and discusses further fields of
application. %as well as relations with other scientific fields.

\section{Occurence Probabilities of Paths}

Let ${\cal M} = \{1,2, \dots,N\}$ be the discrete set of possible states of 
the considered stochastically behaving
system. Moreover, let $w(j|i)$ represent the {\em transition
rate} (i.e. the transition probability per unit time) for state changes
from state $i \in {\cal M}$ to state $j \ne i$. In the following we will
assume that $w(j|i)$ is time-independent. Then, the {\em master equation}
of the corresponding Markovian stochastic process reads
\begin{equation}
 \frac{d}{dt} P(i,t) = \sum_{j=1\atop (j \ne i)}^N \Big[ w(i|j)P(j,t)
 - w(j|i) P(i,t) \Big] \, ,
\label{master}
\end{equation}
where $P(i,t)$ denotes the probability of the system to be in state $i$ at the 
time $t$.
\par
Alternatively, we can consider the associated random walk. Let
$i_0$ be the state of the system at the initial time $t_0$, 
and let $t_l$ denote the {\em transition times} at which the
system changes its state from $i_{l-1}$ to $i_l$. %During the time $t$ with
%$t_{k-1} \le t < t_k$ the system stays in state $i_{k-1}$. 
The corresponding {\em stochastic time series} up to time $t$ with
$t_n \le t < t_{n+1}$ is
\begin{equation}
 (i_0,t_0) \rightarrow (i_1,t_1) \rightarrow \dots \rightarrow (i_n,t_n) 
\end{equation}
and can be numerically generated by means of a {\em Monte-Carlo simulation
method} \cite{Bind}. 
\par
If we are interested in the statistical properties of longitudinal 
(trajectory-related) quantities 
belonging to the considered stochastic process, we need
to evaluate a large number of time series with respect to certain
characteristics. This is connected with a considerable computational effort. 
However, if one is not interested in the respective times $t_l$ at which the
single transitions take place, but only in the {\em path} (sequence of states)
which the system takes, this effort can be very much simplified.
\par
In order to illustrate this, let 
\begin{equation}
 {\cal C}_n := i_0 \rightarrow i_1 \rightarrow \dots \rightarrow i_n
\end{equation}
be the path which the system takes up to the time $t$. 
Note that $w(i_l|i_{l-1}) d\tau_{l-1}$ is the probability of changing from
state $i_{l-1}$ to $i_l$ between time $t_l$ and $t_l + d\tau_{l-1}$ and that
$\mbox{e}^{-w_l\tau_l}$ with the {\em departure rate (overall transition rate)}
\begin{equation}
 w_l := \sum_{i=1\atop (i \ne i_l)}^N w(i|i_l) 
\end{equation} 
is the {\em (survival) probability} 
with which the system stays in state $i_l$ for a time
interval (the {\em survival time})
$\tau_l := t_{l+1} - t_l > 0$ ($\tau_n := t - t_n$). Therefore,
multiplying the probability $P(i_0,t_0)$ of the initial state $i_0$ with
the survival probabilities as well as
the probabilities of the transitions involved and, afterwards,
integrating with respect to the possible survival times $\tau_l$, we obtain the
{\em occurence probability} $P({\cal C}_n,\tau)$ with which
the system takes the path ${\cal C}_n$ up to the time $t := t_0 + \tau$
\cite{Nils}: 
\begin{equation}
 P({\cal C}_n,\tau) = \mbox{e}^{-w_n\tau_n}
 \!\int\limits_0^\infty\! d\tau_{n-1}\, w(i_n|i_{n-1})
 \mbox{e}^{-w_{n-1}\tau_{n-1}} \dots
 \!\int\limits_0^\infty\! d\tau_0 \,   w(i_1|i_0) \mbox{e}^{-w_0\tau_0}
 \delta\!\left( \sum_{l=0}^n \tau_l - \tau \right)\! P(i_0,t_0) \, .
\end{equation}
Here, the $\delta$-function guarantees that the survival times sum up to
the available time $\tau$. Inserting its Laplace-representation 
\begin{equation}
 \delta\left( \sum_{l=0}^n \tau_l - \tau \right) = \frac{1}{2\pi {\rm i}}
 \int\limits_{-c-{\rm i}\infty}^{-c+{\rm i}\infty} du \,
 \exp\left[ u \left( \sum_{l=0}^n \tau_l - \tau \right)\right]
\end{equation}
with a suitable constant $c > 0$ and carrying out the integrations with
respect to the survival times 
$\tau_l$, we finally obtain the formula \cite{Nils,Hel94,Hel95}
\begin{equation}
 P({\cal C}_n,\tau) = 
 \frac{1}{2\pi {\rm i}} \int\limits_{-c-{\rm i}\infty}^{-c+{\rm i}\infty}
 \!\! du \; \frac{\mbox{e}^{-u\tau}}
 {\displaystyle \prod_{l=0}^n (w_l - u)} \, w({\cal C}_n) P(i_0,t_0) \, ,
\label{explicit}
\end{equation}
where we have introduced the abbreviation
\begin{equation}
 w({\cal C}_n) %\equiv w(x_0\rightarrow \dots \rightarrow x_n) 
 := \left\{ 
\begin{array}{ll}
 \delta_{i i_0} & \mbox{if } n=0 \\
 \displaystyle \prod_{l=1}^n w(i_l|i_{l-1}) & \mbox{if } n \ge 1
\end{array}\right.
\label{abk}
\end{equation}
with the Kronecker function $\delta_{ij}$.
\par
It is obviously much easier to calculate formula 
(\ref{explicit}) than to count and evaluate
a huge number of stochastic trajectories. 
However, for this calculation we must first evaluate
the integral by means of the {\em residue theorem}.
If all departure rates $w_l$ are identical (i.e.
$w_l \equiv w$), we get
\begin{equation}
 P({\cal C}_n,\tau) = \frac{\tau^n}{n!} 
 \mbox{e}^{-w\tau}w({\cal C}_n) P(i_0,t_0) \, ,
\label{exp1}
\end{equation}
whereas we obtain
\begin{equation}
 P({\cal C}_n,\tau) = \sum_{k=0}^n f_n(w_k,\tau) w({\cal C}_n)
 P(i_0,t_0) 
\label{exp2}
\end{equation}
with
\begin{equation}
 f_n(u,\tau) := \frac{\mbox{e}^{-u \tau}}
 {\displaystyle \prod_{l=0 \atop (w_l\ne u)}^n (w_l - u)}
\label{fn}
\end{equation}
if the departure rates $w_l$ are pairwise different from each other. However,
serious problems arise in deriving a general result for cases where
{\em different} departure rates $w_l$ occur, but
some of them {\em multiple} times.

\section{Degenerate Paths}

In the following, we will call paths for which {\em some} departure rate $w_l$ 
occurs twice or more often {\em degenerate paths}. Note that all
paths with repeatedly visited states are degenerate! 
\par
For this reason let $m_l \equiv m(w_l)$ 
be the {\em multiplicity} of the departure rate $w_l$ in path ${\cal C}_n$
($l \in \{0,1,\dots ,n\}$). Then we have the relation 
%\begin{equation}
$\sum_{w_l} m_l = n+1$, %\, ,
%\end{equation}
and formula (\ref{explicit}) can be rewritten in the form
\begin{equation}
 P({\cal C}_n,\tau) = -
 \frac{1}{2\pi {\rm i}} \int\limits_{-c+{\rm i}\infty}^{-c-{\rm i}\infty}
 \!\! du \; \frac{\mbox{e}^{-u\tau} w({\cal C}_n)}
 {\displaystyle \prod_{w_l} (w_l - u)^{m_l}} P(i_0,t_0) \, ,
\end{equation}
where we have changed the integration direction. Applying the residue
theorem, we obtain
\begin{eqnarray}
 P({\cal C}_n,\tau) &=& - \sum_{w_k}
 \frac{1}{(m_k - 1)!} \frac{\partial^{m_k - 1}}{\partial u^{m_k - 1}}
 \left( \vphantom{\int\limits_b^b} \right.
 \frac{\mbox{e}^{-u\tau} (-1)^{m_k}}{\displaystyle
 \prod_{w_l\atop (w_l \ne w_k)} (w_l - u)^{m_l}} 
 \left. \left. \vphantom{\int\limits_b^b} \right) \right|_{u=w_k}
 w({\cal C}_n) P(i_0,t_0) \nonumber \\
 &=& \sum_{k=0}^n \frac{(-1)^{m_k + 1}}{ m_k !}
 f_n^{(m_k-1)}(w_k,\tau) w({\cal C}_n) P(i_0,t_0) \, ,
\label{repre}
% \frac{\partial^{n_k - 1}}{\partial u^{n_k - 1}}
% \left. \left[ \frac{\mbox{e}^{-u\tau}}{\displaystyle
% \prod_{l=0\atop (l \ne k)}^n (w_l - u)} \right] \right|_{u=w_k}
% w({\cal C}_n) P(i_0,t_0) \, .
\end{eqnarray}
where $f_n^{(m)}(u,\tau) := \partial^m f_n(u,\tau)/\partial u^m$.
%and $f^{(-1)}(u,\tau) := 0$. 
Introducing the functions
\begin{equation}
 p_{m}(u,\tau) := \frac{(-1)^{m+1}}{f_n(u,\tau) m!} 
 f_n^{(m - 1)}(u,\tau) 
\label{poly}
\end{equation}
we can represent (\ref{repre}) in the form
\begin{equation}
 P({\cal C}_n,\tau) = \sum_{k=0}^n p_{m_k}(w_k,\tau) f_n(w_k,\tau)
 w({\cal C}_n) P(i_0,t_0) \, .
\label{form}
\end{equation}
For $p_{m}(u,\tau)$ we can derive the recursive relation
\begin{eqnarray}
 p_{m}(u,\tau) &=& \frac{-1}{m f_n(u,\tau)} \frac{\partial}
 {\partial u} \left[ f_n(u,\tau) \frac{(-1)^{m}}{f_n(u,\tau) (m-1)!} 
 f_n^{(m-2)}(u,\tau) \right] \nonumber \\
% &=& \frac{-1}{m f_n(u,\tau)} \frac{\partial}{\partial u}
% [ f_n(u,\tau) p_{m-1}(u,\tau) ] \nonumber \\
 &=& \frac{-1}{m} \Bigg[ \Bigg( \sum_{l=0\atop (w_l \ne u)}^n
 \frac{1}{w_l - u} - \tau \Bigg) p_{m-1}(u,\tau) +
 \frac{\partial}{\partial u} p_{m - 1}(u,\tau) \Bigg] \, .
\label{recurs}
\end{eqnarray}
This shows that $p_{m}(u,\tau)$ is a polynomial in $\tau$ of order
$m-1$ since a comparison with (\ref{exp2}) provides us with
%\begin{equation}
$p_1(u,\tau) = 1$. %\, .
%\end{equation}
However, formula (\ref{recurs}) is not very
suitable for a numerical implementation
or a derivation of the explicit form of $p_{m}(u,\tau)$. 
In order to evaluate (\ref{poly}), we try to find
the $(m-1)$st derivative $f_n^{(m-1)}$ of
$f_n$. Since we would not succeed with the usual procedure, we need a trick
and take the `detour' over the first derivative
\begin{equation}
 f_n^{(1)}(u,\tau) = \Bigg( \sum_{l=0\atop (w_l \ne u)}^n
 \frac{1}{w_l - u} - \tau \Bigg) f_n(u,\tau) \, .
\end{equation}
The $(n_0-1)$st derivative of this relation can now be determined as usual:
\begin{eqnarray}
 f_n^{(n_0)}(u,\tau) &=& \sum_{l=0}^{n_0-1}
 { n_0 - 1 \choose l }
 \frac{\partial^l}{\partial u^l} \Bigg( \sum_{l=0\atop (w_l \ne u)}^n
 \frac{1}{w_l - u} - \tau \Bigg) \frac{\partial^{n_0-1-l}}{\partial 
 u^{n_0-1-l}} f_n(u,\tau) \nonumber \\
 &=& \sum_{n_1 = 0}^{n_0-1} \frac{(n_0 - 1)!}{n_1 !}
 g^{(n_0 - n_1)}(u,\tau)f_n^{(n_1)}(u,\tau) \, .
\label{recursiv}
\end{eqnarray}
Here, we have used the definitions $n_1 := n_0 - 1 - l$ and
\begin{equation}
 g^{(l+1)}(u,\tau) := \frac{1}{l !}
 \frac{\partial^l}{\partial u^l} \Bigg( \sum_{l=0\atop (w_l \ne u)}^n
 \frac{1}{w_l - u} - \tau \Bigg)
 = \sum_{l=0\atop(w_l \ne u)}^n \frac{1}{(w_l - u)^{l+1}} 
 - \tau \delta_{l0} \, .
\end{equation}
Formula (\ref{recursiv}) obviously allows to express the derivatives
$f_n^{(n_0)}(u,\tau)$ in terms of derivatives $f_n^{(n_1)}(u,\tau)$
of lower order $n_1 < n_0$. 
Therefore, we insert for $f_n^{(n_1)}(u,\tau)$ again
relation (\ref{recursiv}), etc. After a couple of iterations this 
procedure leads, for $n_0 \ge 1$, to
\begin{eqnarray}
  f_n^{(n_0)} &=& (n_0 - 1)! 
  \Bigg( g^{(n_0)} f_n^{(0)} + \sum_{n_1 = 1}^{n_0 - 1}
 \frac{g^{(n_0-n_1)} }{n_1} \Bigg( g^{(n_1)} f_n^{(0)} 
 + \sum_{n_2 = 1}^{n_1 - 1} \frac{g^{(n_1 - n_2)}}{n_2} \nonumber \\
 &\times & \Bigg( g^{(n_2)} f_n^{(0)} + \dots
 \Bigg( g^{(n_{n_0 - 1})} f_n^{(0)}
 + \sum_{n_{n_0} = 1}^{n_{n_0 - 1} - 1}
 \frac{g^{(n_{n_0 - 1} - n_{n_0})}}{n_{n_0} !} f_n^{(n_{n_0})}
 \Bigg) \dots \Bigg) \Bigg) \Bigg) \, ,
\label{in}
\end{eqnarray}
where we have separated the terms containing $f_n^{(0)}(u,\tau) = f_n(u,\tau)$
from the rest and omitted the arguments $(u,\tau)$ of the functions.
In equation (\ref{in}) we have to apply the convention $\sum_{n=l_1}^{l_2}
(...) := 0$ if $l_2 < l_1$ (as will be shown by Fig. 1). 
Consequently, the term with the $n_{n_0}$th
derivative $f_n^{(n_{n_0})}$ does not give a contribution to
$f_n^{(n_0)}$, since $\max(n_1) = n_0 - 1$, $\max(n_2) = n_1 - 1
= n_0 - 2$, $\dots$, $\max(n_{n_0}) = n_0 - n_0 = 0$. Therefore, we obtain
the final result
\begin{eqnarray}
 p_{m} &=& \frac{(-1)^{m+1}}{m (m - 1)}
 \Bigg( g^{(m-1)} + \sum_{n_1 = 1}^{m - 2}
 \frac{g^{(m-1-n_1)}}{n_1} \Bigg( g^{(n_1)} 
 + \sum_{n_2 = 1}^{n_1 - 1} \frac{g^{(n_1 - n_2)}}{n_2} \nonumber \\
&\times & \Bigg( \dots + \sum_{n_{n_0} -1 = 1}^{n_{n_0 - 2} - 1}
 \frac{g^{(n_{n_0-2} - n_{n_0 - 1})}}{n_{n_0-1}}
 \Bigg( g^{(n_{n_0 - 1})} \Bigg) \dots \Bigg) \Bigg) \Bigg) 
\label{endlich}
\end{eqnarray}
for $m \ge 2$. The correctness of this formula is illustrated by 
Fig. 1.
\marginpar[]{Fig. 1}
\section{Numerical Evaluation of the `Contracted Path-Integral Solution'}

In Refs. \cite{Hel94,Hel95} it has been proved that the solution $P(i,t)$ of master
equation (\ref{master}) can be represented in the form
\begin{equation}
 P(i,t_0 + \tau) = \sum_{n=0}^\infty \sum_{{\cal C}_n} P({\cal C}_n,\tau)
 := \sum_{n=0}^\infty \sum_{i_{n-1}=1\atop (i_{n-1} \ne i_n)}^N
 \sum_{i_{n-2}=1\atop (i_{n-2} \ne i_{n-1})}^N \dots
 \sum_{i_0=1\atop (i_0 \ne i_1)}^N P(i_0\rightarrow i_1 \rightarrow
 \dots \rightarrow i_n,\tau)
\label{pathint}
\end{equation}
with $i_n := i$. According to (\ref{pathint}), the probability 
$P(i,t_0 + \tau)$ to find state
$i$ at the time $t = t_0 + \tau$ is given as the sum over the occurence
probabilities $P({\cal C}_n,\tau)$
of all paths ${\cal C}_n$ which have an arbitrary length
$n$ but have led to state $i_n = i$ within the time interval $\tau$.
\par
When (\ref{pathint}) is evaluated numerically, one must 
restrict the summation to a finite number of {\em relevant} paths.
Here, we will define a path ${\cal C}_n$ to be relevant 
if it fulfils the condition
\begin{equation}
 | \tau - \langle \tau \rangle_{{\cal C}_n} | \le a \sqrt{\theta_{{\cal C}_n}}
 \, ,
\label{cond}
\end{equation}
where 
%\begin{equation}
$\langle \tau \rangle_{{\cal C}_n} := \sum_{k=0}^n 1/w_k$
%\end{equation}
is the {\em mean value} of the {\em occurence times} of path ${\cal C}_n$ and
%\begin{equation}
$\theta_{{\cal C}_n} := \sum_{k=0}^n 1/(w_k)^2$
%\end{equation}
their {\em variance} (cf. \cite{Hel94,Hel95}). 
The parameter $a$ is a measure for the {\em accuracy} 
of the above outlined approximation. According to our experience, $a 
\lesssim 3$ usually guarantees a reconstruction of $99\%$ of the
probability distribution $P(i,t)$ which can be
checked by means of the {\em normalization condition}
%\begin{equation}
$\sum_{i=1}^N P(i,t) = 1$. %\, .
%\end{equation}
%In contrast to the {\em `breadth-first'} procedure \cite{Dep},
%which has been suggested in Ref. \cite{Hel94}, 
\par
An {\em efficient} implementation 
%(with respect to computer time and memory) 
of the approximate 
`contracted path-integral solution' on a serial computer bases on the
path-search procedure {\em `depth-first'} \cite{Dep}. After each
step of this standard procedure 
which extends the previous path ${\cal C}_n$ by an additional state
$i_{n+1}$ (leading to the path ${\cal C}_{n+1} := {\cal C}_n \rightarrow
i_{n+1}$) the following quantities are calculated and stored
in a so-called {\em `linked list'} \cite{Dep}:
\begin{equation}
 \langle \tau \rangle_{{\cal C}_{n+1}} := \langle \tau \rangle_{{\cal C}_n}
 + \frac{1}{w_{n+1}} \, , \qquad
 \theta_{{\cal C}_{n+1}} := \theta_{{\cal C}_n} + \frac{1}{(w_{n+1})^2} \, ,
 \qquad
 w({\cal C}_{n+1}) := w({\cal C}_n) w(i_{n+1}|i_n) \, , 
\end{equation}
and
\begin{equation}
 f_{n+1}(w_k,\tau) := \frac{f_n(w_k,\tau)}{w_{n+1} - w_k} \quad \mbox{for}
 \quad k=0,\dots,n \quad \mbox{with} \quad w_k \ne w_{n+1} \, .
\end{equation}
If condition (\ref{cond}) is fulfilled, the occurence probability
$P({\cal C}_n,\tau)$ is calculated according to formula (\ref{form})
and added to $P(i_{n+1},\tau)$. Afterwards, the next step of the
`depth-first' procedure is carried out. 
\par
The path ${\cal C}_{n+1}$ is not further extended if 
%\begin{equation}
$\langle \tau \rangle_{{\cal C}_{n+1}} - a \sqrt{\theta_{{\cal C}_{n+1}}}
 > \tau$.
%\end{equation}
This implies that the path ${\cal C}_{n+1}$ and all longer paths 
${\cal C}_{n+1} \rightarrow i_{n+2} \rightarrow \dots$
which include ${\cal C}_{n+1}$ as subpath are irrelevant. Then, the procedure
traces back its path ${\cal C}_{n+1}$ one step before it tries to extend
the previous path ${\cal C}_n$ by another state $i'_{n+1} = i_{n+1} + 1$.
\par
Figure 2 illustrates the approximate `contracted path-integral
solution' for the example of {\em Brownian motion} which is characterized by
nearest-neighbor transitions with the following transition rates:
\begin{equation}
 w(j|i) = \left\{ \begin{array}{ll}
D & \mbox{if } |j-i| = 1 \\
0 & \mbox{otherwise.}
\end{array} \right.
\end{equation}
\marginpar[]{Fig. 2}
\section{Summary and Fields of Application}

In this paper we were able to derive 
an analytical formula for the occurence probability
of arbitrary paths including the complex case of repeatedly visited
states and/or equal
departure rates. This allows a numerical implementation of the
`contracted path-integral solution' of the discrete master equation. 
Other kinds of path-integral formalisms have been developed for
{\em chaotic mappings} \cite{Tel},
the {\em Schr\"odinger equation} \cite{schr},
the {\em Fokker-Planck equation} \cite{Risk,Hakpath}, 
and also the {\em master equation} \cite{Hakpath}.
\par
The importance of the occurence-probability formula 
goes far beyond the new solution method for the master equation.
It considerably simplifies the evaluation of the trajectories related 
with random walk models. Therefore, a simulation program for the calculation
of the occurence probabilities
of paths, the path-integral solution, most probable paths, and first-passage
times has recently been developed at the University of Stuttgart.
It is expected to be a useful tool for investigations in a number of current
research fields concerning 
different types of {\em random walks} \cite{random},
{\em noise-induced transitions} \cite{noise},
{\em first-passage time problems} \cite{passage},
{\em percolation} \cite{percol},
{\em critical behavior} \cite{critical},
and {\em diffusion in disordered or fractal media} \cite{diffusion}.
%and {\em crack propagation} \cite{crack}.

\clearpage
\unitlength1.5cm
\begin{figure}[htbp]
\begin{center}
\begin{picture}(7,6)
\put(0.8,0.3){\epsfig{height=8\unitlength, angle=0,
      bbllx=201pt, bblly=50pt, bburx=554pt, bbury=770pt,
      file=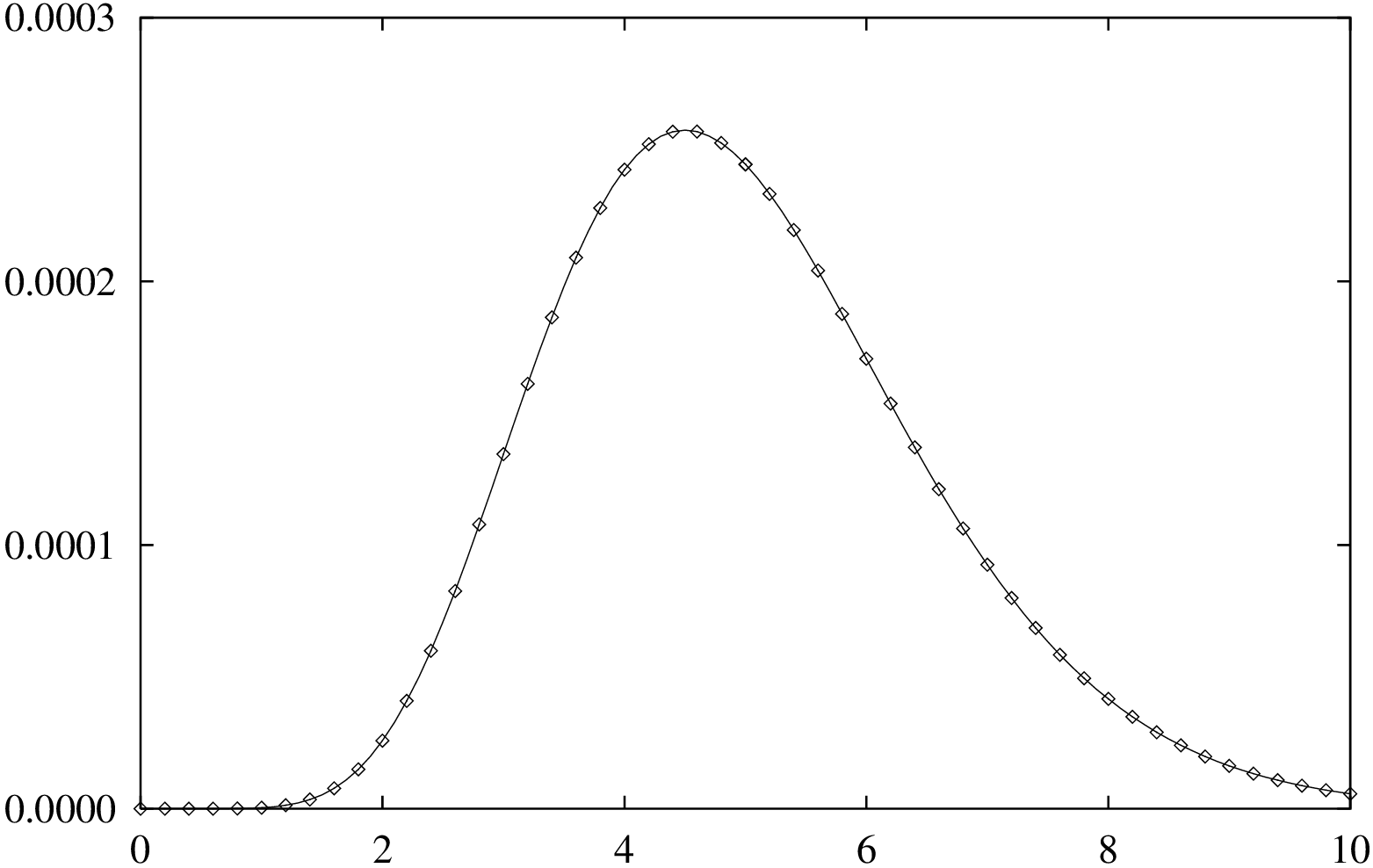}}
\put(3.5,0){\makebox(0,0){$\tau$}}
\put(0,5.5){\makebox(0,0){$\displaystyle \frac{P({\cal C}_n,\tau)}
{w({\cal C}_n)P(i_0,t_0)}$}}
\end{picture}
\end{center}
\caption[]{Comparison of the occurence probability $P({\cal C}_n,\tau)$
calculated via formulas (\ref{form}), (\ref{fn}) and (\ref{endlich}) 
for the case $w_l \equiv w$, $m_l = n+1 = 10$ ($\diamond$) 
with the occurence probability of the same path ${\cal C}_n$
according to (\ref{exp1}) (---).} 
\end{figure}
\par
\begin{figure}[htbp]
\begin{center}
\begin{picture}(7,5)
\put(1.5,0){\epsfig{height=8\unitlength, angle=0,
      bbllx=201pt, bblly=50pt, bburx=554pt, bbury=770pt,
      file=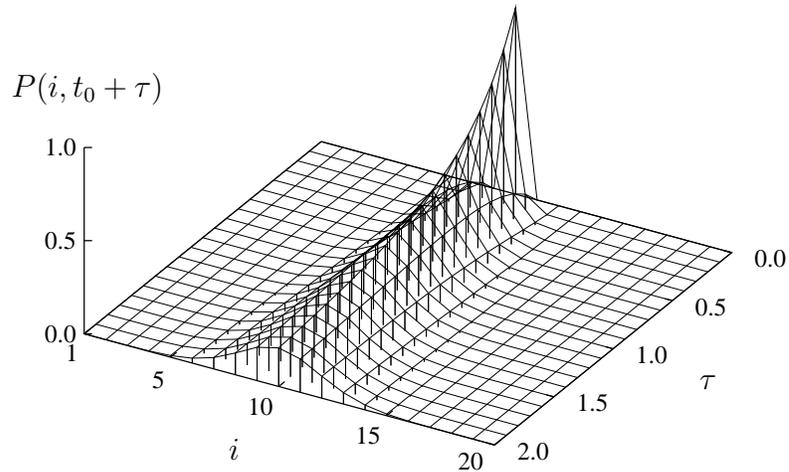}}
\put(2.3,0.6){\makebox(0,0){$i$}}
\put(6.5,1.2){\makebox(0,0){$\tau$}}
\put(1,3.8){\makebox(0,0){$P(i,t_0+\tau)$}}
\end{picture}
\end{center}
\caption[]{Comparison of the approximate path-integral solution for Brownian 
motion (bars) with the numerical result of Runge-Kutta
integration \cite{Rung} for the corresponding master equation (grid).
The diffusion coefficient was assumed to be $D=1$, and the accuracy 
parameter was set to $a=3$ so that the path-integral method reconstructs
about 99\% of the exact probability distribution.}
\end{figure}

\end{document}